\documentclass[twocolumn,showpacs,preprintnumbers]{revtex4}
\usepackage{graphicx}
\usepackage{dcolumn}
\usepackage{bm}

\begin{document}

\title{Effective Hamiltonian for fermions in an optical lattice across Feshbach
resonance}
\author{L.-M. Duan}

\address{FOCUS center and MCTP, Department of Physics,
University of Michigan, Ann Arbor, MI 48109 }

\begin{abstract}
We derive the Hamiltonian for cold fermionic atoms in an optical
lattice across a broad Feshbach resonance, taking into account of
both multiband occupations and neighboring-site collisions. Under
typical configurations, the resulting Hamiltonian can be
dramatically simplified to an effective single-band model, which
describes a new type of resonance between the local dressed
molecules and the valence bond states of fermionic atoms at
neighboring sites. On different sides of such a resonance, the
effective Hamiltonian is reduced to either a $t$-$J$ model for the
fermionic atoms or an XXZ model for the dressed molecules. The
parameters in these models are experimentally tunable in the full
range, which allows for observation of various phase transitions.
\end{abstract}

\maketitle

Recently, there are many exciting advances in the ultracold atoms physics
\cite{1,2}. For these developments, two experimental control techniques play
the critical role: one is the Feshbach resonance to control the interaction
magnitude between the atoms \cite{3}, and the other is the optical lattice
to introduce diverse interaction configurations \cite{4}. It is natural to
consider combination of these two techniques, and indeed, significant
efforts have been put forward towards this direction \cite{5,6,7,8,9}.

A fundamental problem along this direction is to derive an
appropriate Hamiltonian for this strongly interacting system which
can serve as the starting point for further investigations. A
number of generalizations of the Hubbard model have been proposed
to describe this system by including the on-site atom-molecule
coupling, typically ignoring the upper-band occupations \cite{5,6}
(see the comment in Ref. \cite{7}). These generalizations may
model physics associated with a very narrow resonance, however,
they are not adequate to describe typical broad Feshbach
resonance, such as for $^{40}K$ or $^{6}Li$. For the latter case,
first, one needs to include all the multiband coupling terms even
if the system temperature is well below the band gap. The reason
is that the strong atom-molecule coupling results in population of
the upper bands (the on-site coupling rate is typically larger
than the band gap) \cite{7}. Second, one also needs to include the
atom-molecule coupling from the neighboring sites which has been
ignored in all the previous works. We will see that off-site
atom-molecule coupling is typically larger than the atom
tunnelling rate, and inclusion of these off-site interactions
leads to qualitatively different physics.

In this paper we rigorously derive the interaction Hamiltonian for fermionic
atoms in an optical lattice across a broad Feshbach resonance, taking into
account of both multiband couplings and off-site interactions. The strong
on-site interaction between the atoms make them first form local dressed
molecules. Under typical experimental conditions, we then derive an
effective single-band Hamiltonian, describing the resonant interaction
between the local dressed molecules and the valence bonds (singlets) of
fermions at neighboring sites \cite{10}. In this effective single-band
resonance model (different from the conventional Feshbach resonance),
multi-band occupations have been incorporated through the local dressed
molecule states. On different sides of this resonance, the effective
Hamiltonian is reduced to either a $t$-$J$ model for the fermionic atoms or
an XXZ (anisotropic Heisenberg) model for the dressed molecules, opening up
the prospect of using this system to probe some fundamental physics
associated with the latter two models.

We consider fermionic atoms with two internal states labelled by
the spin index $\sigma =\uparrow ,\downarrow $. The atoms are
loaded into an optical lattice, and tuned close to a Feshbach
resonance by an external magnetic field. The Hamiltonian, in terms
of the field operators $\Psi ^{\left( m\right) }\left(
\mathbf{r}\right) ,\Psi _{\sigma }^{\left( a\right) }\left(
\mathbf{r}\right) $ respectively for the bare molecules and the
fermionic atoms, then has the form $H=H_{0}+H_{I},$ with
$H_{0}=\sum_{\sigma =\uparrow ,\downarrow }\int \Psi _{\sigma
}^{\left( a\right) \dagger }\left( T_{a}+V_{a}\right) \Psi
_{\sigma }^{\left( a\right) }d^{3}\mathbf{r}+\int \Psi ^{\left(
m\right) \dagger }\left( T_{m}+V_{m}+\nu _{b}\right) \Psi ^{\left(
m\right) }d^{3}\mathbf{r}$ and $H_{I}=\left( \alpha \int \Psi
^{\left( m\right) \dagger }\Psi _{\uparrow }^{\left( a\right)
}\Psi
_{\downarrow }^{\left( a\right) }d^{3}\mathbf{r+}h.c.\right) \mathbf{+}%
U_{bg}\int \Psi _{\downarrow }^{\left( a\right) \dagger }\Psi
_{\uparrow }^{\left( a\right) \dagger }\Psi _{\uparrow }^{\left(
a\right) }\Psi _{\downarrow }^{\left( a\right) }d^{3}\mathbf{r.}$
In the above expression, the kinetic energy $T_{a}=2T_{m}=-\hbar
^{2}\nabla ^{2}/2m$ ($m$ is the atom mass), and the potential
energy $V_{a}=V_{m}/2=V_{0}\left[ \sin ^{2}k_{0}x+\sin
^{2}k_{0}y+\sin ^{2}k_{0}z\right] .$ The $V_{a}$ and $V_{m}$ are
due to the optical lattice potential from far-off-resonant laser
beams with a wave vector $k_{0}=2\pi /\lambda $. For simplicity of
the notation, the potential depth $V_{0}$ is assumed to be the
same along the $x,y,z$ directions. The detuning $\nu _{b}$ of the
bare molecules can be controlled by an external magnetic field
$B$. The atom-molecule coupling rate $\alpha $ and the atom
background scattering rate $U_{bg}$ are determined from the atom
scattering length as $\alpha =\sqrt{4\pi \hbar ^{2}\mu
_{co}W\left| a_{b}\right| /m}$, $U_{bg}=4\pi \hbar ^{2}a_{b}/m$,
where we have assumed
the atom scattering length near Feshbach resonance takes the form $%
a_{s}=a_{b}\left( 1-W/\left( B-B_{0}\right) \right) $, with $a_{b}$, the
background scattering length, $B_{0}$, the resonance point, $W$, the
resonance width, and $\mu _{co}$, the difference of the atom magnetic
moments between the closed and the open scattering channels.

The filed operators $\Psi _{\sigma }^{\left( a\right) }$ and $\Psi ^{\left(
m\right) }$ can be expanded with the Wannier functions associated with the
lattice potential in the forms $\Psi _{\sigma }^{\left( a\right) }\left(
\mathbf{r}\right) =\sum_{i\mathbf{p}}a_{i\mathbf{p\sigma }}w_{\mathbf{p}%
}^{\left( a\right) }\left( \mathbf{r-r}_{i}\right) $, $\Psi ^{\left(
m\right) }\left( \mathbf{r}\right) =\sum_{i\mathbf{p}}b_{i\mathbf{p}}w_{%
\mathbf{p}}^{\left( m\right) }\left( \mathbf{r-r}_{i}\right) $, where $w_{%
\mathbf{p}}^{\left( a\right) }\left( \mathbf{r-r}_{i}\right) \equiv
w_{p_{x}}^{\left( a\right) }\left( x-x_{i}\right) w_{p_{y}}^{\left( a\right)
}\left( y-y_{i}\right) w_{p_{z}}^{\left( a\right) }\left( z-z_{i}\right) $ ($%
w_{\mathbf{p}}^{\left( m\right) }\left( \mathbf{r-r}_{i}\right) $) are the
Wannier functions for atoms (molecules) at the site $\mathbf{r}_{i}$ with $%
\mathbf{p}\equiv \left( p_{x},p_{y},p_{z}\right) $ labelling different
lattice bands, and $a_{i\mathbf{p\sigma }}$ ($b_{i\mathbf{p}}$) are the
associated mode operators. With these expansions, the Hamiltonian $H_{0}$
then has the form $H_{0}=\sum_{i\mathbf{p}}\left( \epsilon _{\mathbf{p}%
}^{\left( m\right) }+\nu _{b}\right) b_{i\mathbf{p}}^{\dagger }b_{i\mathbf{p}%
}+\sum_{i\mathbf{p\sigma }}\epsilon _{\mathbf{p}}^{\left( a\right) }a_{i%
\mathbf{p}\sigma }^{\dagger }a_{i\mathbf{p}\sigma }+\sum_{i\mathbf{p}%
}\sum_{j\in N\left( i\right) }\left( t_{\mathbf{p}}^{\left( m\right) }b_{i%
\mathbf{p}}^{\dagger }b_{i\mathbf{p}}+\sum_{\mathbf{\sigma }}t_{\mathbf{p}%
}^{\left( a\right) }a_{i\mathbf{p}\sigma }^{\dagger
}a_{j\mathbf{p}\sigma }\right) $ up to the nearest-neighbor
tunnelling, where $N\left( i\right) $ denotes the neighboring
sites of $i$, $\epsilon _{\mathbf{p}}^{\left( m\right) }=\epsilon
_{\mathbf{p}}^{\left( a\right) }\approx \left[
2(p_{x}+p_{y}+p_{z})+3\right] \sqrt{V_{0}E_{r}}$ under the
harmonic approximation to the potential well ($E_{r}\equiv \hbar
^{2}k_{0}^{2}/2m$ is the atom recoil energy), and the tunnelling
rates $t_{\mathbf{p}}^{\left( m\right) },t_{\mathbf{p}}^{\left(
a\right) }$ can be determined through the standard band
calculation (see Fig. 1 and Ref. \cite{13}).

In the expansion for $H_{I}$, usually one ignores all the terms except the
ones for the on-site interaction. However, for typical broad Feshbach
resonance, the nearest-neighbor atom-molecule coupling rates $c_{1;\mathbf{%
pss}^{\prime }}^{\left( am\right) }\equiv \alpha \int w_{\mathbf{p}}^{\left(
m\right) \ast }\left( \mathbf{r}\right) w_{\mathbf{s}}^{\left( a\right)
}\left( \mathbf{r}\right) w_{\mathbf{s}^{\prime }}^{\left( a\right) }\left(
\mathbf{r-\delta r}\right) d^{3}\mathbf{r}$ ($\left| \mathbf{\delta r}%
\right| \equiv \left| \mathbf{r}_{j}-\mathbf{r}_{i}\right| $ is the lattice
constant) can be significantly larger than the atom tunnelling rates $t_{%
\mathbf{p}}^{\left( a\right) }$, and thus should not be neglected. To see
that, we have numerically calculated the exact Wannier functions and their
overlaps, and some of the results are shown in Fig. 1. The calculation
clearly shows that the neighboring couplings are not negligible. For
instance, for the lowest band with $\mathbf{p=s=s}^{\prime }\mathbf{=}\left(
0,0,0\right) $, the rates $c_{1;\mathbf{pss}^{\prime }}^{\left( am\right) }$
and $t_{\mathbf{p}}^{\left( a\right) }$ scale with the potential depth $%
V_{0}/E_{r}$ by roughly the same exponential form, and the ratios between
them are estimated to be $c_{1;\mathbf{pqs}}^{\left( am\right) }/t_{\mathbf{p%
}}^{\left( a\right) }\approx 10\sqrt{V_{0}/E_{r}}$ \ ($1.4\sqrt{V_{0}/E_{r}}$%
)\ respectively for $^{6}Li$ ($^{40}K$) (see Fig. 1). For this estimation,
we have taken the following parameters for $^{6}Li$ ($^{40}K$): $W=180G$ ($%
8G $), $a_{b}=-2000a_{B}$ ($170a_{B}$), $\lambda =1\mu m$ ($0.8\mu m$), $\mu
_{co}\approx 2\mu _{B}$ ($\mu _{B}$: Bohr magneton; $a_{B}$: Bohr radius).
So, we include here all the nearest neighbor coupling terms in the expansion
for $H_{I}$. The next nearest neighbor couplings can be safely neglected
unless the lattice potential is very weak with $V_{0}/E_{r}<3$. Up to this
order, the interaction Hamiltonian $H_{I}$ is expressed as
\begin{widetext}

\begin{eqnarray}
H_{I} &=&\sum_{i\mathbf{pqs}}\left( c_{0;\mathbf{pss}^{\prime }}^{\left(
am\right) }b_{i\mathbf{p}}^{\dagger }a_{i\mathbf{s}\uparrow }a_{i\mathbf{s}%
^{\prime }\downarrow }+h.c.\right) +\sum_{i\mathbf{pqss}^{\prime }}c_{0;%
\mathbf{pqss}^{\prime }}^{\left( aa\right) }a_{i\mathbf{p}\downarrow
}^{\dagger }a_{i\mathbf{q}\uparrow }^{\dagger }a_{i\mathbf{s}\uparrow }a_{i%
\mathbf{s}^{\prime }\downarrow }  \nonumber \\
&&+\sum_{i}\sum_{j\in N\left( i\right) }\left[ \left( \sum_{\mathbf{pss}%
^{\prime }}c_{1;\mathbf{pss}^{\prime }}^{\left( am\right) }b_{i\mathbf{p}%
}^{\dagger }+\sum_{\mathbf{pqss}^{\prime }}c_{1;\mathbf{pqss}^{\prime
}}^{\left( aa\right) }a_{i\mathbf{p\downarrow }}^{\dagger }a_{i\mathbf{%
q\uparrow }}^{\dagger }\right) \left( a_{i\mathbf{s}\uparrow }a_{j\mathbf{s}%
^{\prime }\downarrow }-a_{i\mathbf{s}\downarrow }a_{j\mathbf{s}^{\prime
}\uparrow }\right) +\sum_{\mathbf{pss}^{\prime }}c_{2;\mathbf{pss}^{\prime
}}^{\left( am\right) }b_{i\mathbf{p}}^{\dagger }a_{j\mathbf{s}\uparrow }a_{j%
\mathbf{s}^{\prime }\downarrow }+h.c.\right]   \nonumber \\
&&+\sum_{i\mathbf{pqss}^{\prime }}\sum_{j\in N\left( i\right) }\left[ c_{2;%
\mathbf{pqss}^{\prime }}^{\left( aa\right) }a_{i\mathbf{p\downarrow }%
}^{\dagger }a_{i\mathbf{q\uparrow }}^{\dagger }a_{j\mathbf{s}\uparrow }a_{j%
\mathbf{s}^{\prime }\downarrow }+c_{3;\mathbf{pqss}^{\prime }}^{\left(
aa\right) }a_{i\mathbf{p\downarrow }}^{\dagger }a_{j\mathbf{q\uparrow }%
}^{\dagger }\left( a_{i\mathbf{s}\uparrow }a_{j\mathbf{s}^{\prime
}\downarrow }-a_{i\mathbf{s}\downarrow }a_{j\mathbf{s}^{\prime }\uparrow
}\right) \right] ,
\end{eqnarray}
where $c_{0;\mathbf{pss}^{\prime }}^{\left( am\right)
}\equiv \alpha \int w_{\mathbf{p}}^{\left( m\right) \ast }\left( \mathbf{r}%
\right) w_{\mathbf{s}}^{\left( a\right) }\left( \mathbf{r}\right) w_{\mathbf{%
s}^{\prime }}^{\left( a\right) }\left( \mathbf{r}\right) d^{3}\mathbf{r}$, $%
c_{2;\mathbf{pss}^{\prime }}^{\left( am\right) }\equiv \alpha \int w_{%
\mathbf{p}}^{\left( m\right) \ast }\left( \mathbf{r+\delta r}\right) w_{%
\mathbf{s}}^{\left( a\right) }\left( \mathbf{r}\right) w_{\mathbf{s}^{\prime
}}^{\left( a\right) }\left( \mathbf{r}\right) d^{3}\mathbf{r}$, $c_{0;%
\mathbf{pqss}^{\prime }}^{\left( aa\right) }\equiv U_{bg}\int w_{\mathbf{p}%
}^{\left( a\right) \ast }\left( \mathbf{r}\right) w_{\mathbf{q}}^{\left(
a\right) \ast }\left( \mathbf{r}\right) w_{\mathbf{s}}^{\left( a\right)
}\left( \mathbf{r}\right) w_{\mathbf{s}^{\prime }}^{\left( a\right) }\left(
\mathbf{r}\right) d^{3}\mathbf{r}$, $c_{1;\mathbf{pqss}^{\prime }}^{\left(
aa\right) }\equiv U_{bg}\int w_{\mathbf{p}}^{\left( a\right) \ast }\left(
\mathbf{r}\right) w_{\mathbf{q}}^{\left( a\right) \ast }\left( \mathbf{r}%
\right) w_{\mathbf{s}}^{\left( a\right) }\left( \mathbf{r}\right) w_{\mathbf{%
s}^{\prime }}^{\left( a\right) }\left( \mathbf{r-\delta r}\right) d^{3}%
\mathbf{r}$, $c_{2;\mathbf{pqss}^{\prime }}^{\left( aa\right) }\equiv
U_{bg}\int w_{\mathbf{p}}^{\left( a\right) \ast }\left( \mathbf{r}\right) w_{%
\mathbf{q}}^{\left( a\right) \ast }\left( \mathbf{r}\right) w_{\mathbf{s}%
}^{\left( a\right) }\left( \mathbf{r-\delta r}\right) w_{\mathbf{s}^{\prime
}}^{\left( a\right) }\left( \mathbf{r-\delta r}\right) d^{3}\mathbf{r}$, $%
c_{3;\mathbf{pqss}^{\prime }}^{\left( aa\right) }\equiv \left( U_{bg}\right)
\int w_{\mathbf{p}}^{\left( a\right) \ast }\left( \mathbf{r}\right) w_{%
\mathbf{q}}^{\left( a\right) \ast }\left( \mathbf{r-\delta r}\right) w_{%
\mathbf{s}}^{\left( a\right) }\left( \mathbf{r}\right) w_{\mathbf{s}^{\prime
}}^{\left( a\right) }\left( \mathbf{r-\delta r}\right) d^{3}\mathbf{r}$.
\end{widetext}

\begin{figure}[tbph]
\centering
\includegraphics[height=5cm,width=8cm]{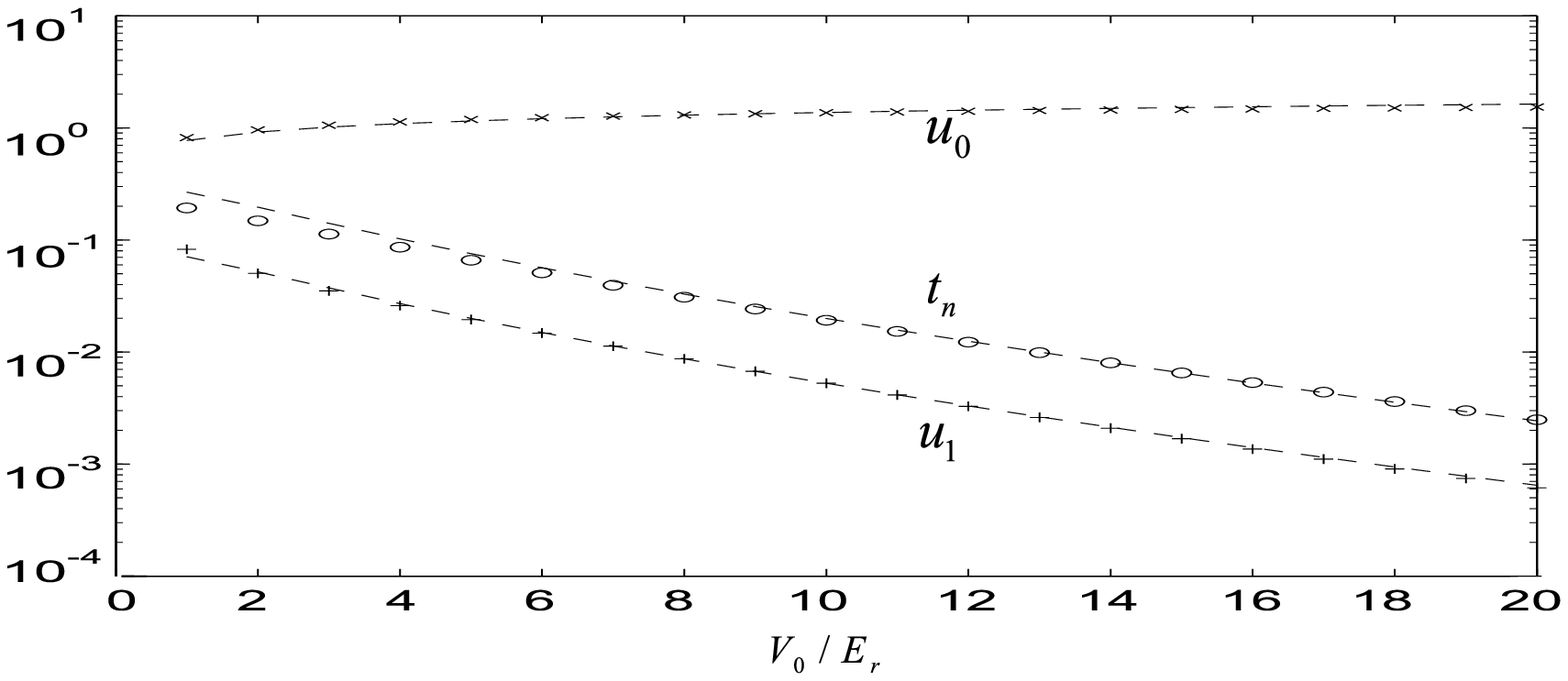}
\caption{ The normalized atom tunnelling rate $t_{n}\equiv t_{\mathbf{0}%
}^{\left( a\right) }/E_{r}$ (the subscript $\mathbf{0}$ means the
lowest band), and the normalized overlaps between the atomic and
the molecular Wannier functions $u_{0}\equiv
\protect\sqrt{\protect\lambda /2}\int \left( w_{0}^{\left(
m\right) }\left( x\right) \right) ^{\ast }\left( w_{0}^{\left(
a\right) }\left( x\right) \right) ^{2}dx$\ (on-site) and
$u_{1}\equiv \protect\sqrt{\protect\lambda /2}\int \left(
w_{0}^{\left( m\right) }\left( x\right) \right) ^{\ast
}w_{0}^{\left( a\right) }\left( x\right) w_{0}^{\left( a\right)
}\left( x+\protect\lambda /2\right) dx$ (neighboring sites) shown
as a function of the lattice depth $V_{0}$ in the unit of the atom
recoil energy $E_{r}$. The ``o'',``$\times $'',``+'' points denote
the exact results from the numerical calculation, and the dashed
curves show
the fits from the formula $t_{n}\simeq \left( 3.5/\protect\sqrt{\protect\pi }%
\right) \left( V_{0}/E_{r}\right) ^{3/4}\exp \left( -2\protect\sqrt{%
V_{0}/E_{r}}\right) $, $u_{0}\simeq 0.77\left( V_{0}/E_{r}\right) ^{1/4}$,
$u_{1}\simeq 0.52\left( V_{0}/E_{r}\right) ^{3/4}\exp \left( -2\protect%
\sqrt{V_{0}/E_{r}}\right) $.}
\label{Fig1}
\end{figure}

The above Hamiltonian seems extremely complicated, however, we show now that
under typical experimental conditions it can be dramatically simplified. To
make this possible, first we note there are several different energy scales
in this system, including the on-site interaction energy $E_{on}\sim c_{0;%
\mathbf{pss}^{\prime }}^{\left( am\right) }$, the band gap $E_{bg}\sim 2%
\sqrt{V_{0}E_{r}}$, the off-site interaction energy $E_{off}\sim c_{1;%
\mathbf{pss}^{\prime }}^{\left( am\right) }$, and the atom tunnelling rate $%
E_{t}\sim t_{\mathbf{p}}^{\left( a\right) }$. In experiments, the energy
scale set by $E_{on},E_{bg}$ is typically much higher than the one set by $%
E_{off},E_{t}$. So we take the approximation $\left( E_{on},E_{bg}\right)
\gg \left( E_{off},E_{t}\right) $. In this case, we can first solve the
single-site Hamiltonian $H_{i}=\sum_{\mathbf{p}}\left[ \left( \epsilon _{%
\mathbf{p}}^{\left( m\right) }+\nu _{b}\right) b_{i\mathbf{p}}^{\dagger }b_{i%
\mathbf{p}}+\sum_{\mathbf{\sigma }}\epsilon _{\mathbf{p}}^{\left( a\right)
}a_{i\mathbf{p}\sigma }^{\dagger }a_{i\mathbf{p}\sigma }\right] +\sum_{%
\mathbf{pqs}}\left( c_{0;\mathbf{pss}^{\prime }}^{\left( am\right) }b_{i%
\mathbf{p}}^{\dagger }a_{i\mathbf{s}\uparrow }a_{i\mathbf{s}^{\prime
}\downarrow }+h.c.\right) +\sum_{\mathbf{pqss}^{\prime }}c_{0;\mathbf{pqss}%
^{\prime }}^{\left( aa\right) }a_{i\mathbf{p}\downarrow }^{\dagger }a_{i%
\mathbf{q}\uparrow }^{\dagger }a_{i\mathbf{s}\uparrow }a_{i\mathbf{s}%
^{\prime }\downarrow }$ on the site $i$. We assume that the average atom
filling number $\overline{n}$ of the lattice is smaller than $2$. If a site
has a single atom on the $p$th band, its energy is just given by $\epsilon _{%
\mathbf{p}}^{\left( a\right) }$. If two atoms occupy the same site, they
will form a local dressed molecule state, which in general can be written in
the form $\left| \Psi _{i\mu }\right\rangle =d_{i\mu }^{\dagger }\left|
0\right\rangle $, where $\left| 0\right\rangle $ denotes the vacuum state
and the dressed molecule creation operator $d_{i\mu }^{\dagger }\equiv \sum_{%
\mathbf{p}}\chi _{\mu \mathbf{p}}b_{i\mathbf{p}}^{\dagger }+\sum_{\mathbf{pq}%
}\gamma _{\mu \mathbf{pq}}a_{i\mathbf{p}\downarrow }^{\dagger }a_{i\mathbf{q}%
\uparrow }^{\dagger }$. In this expression for $d_{i\mu }^{\dagger }$, the
superposition coefficients $\chi _{\mu \mathbf{p}},\gamma _{\mu \mathbf{pq}}$%
, with the normalization $\sum_{\mathbf{p}}\chi _{\mu \mathbf{p}}^{\ast
}\chi _{\mu ^{\prime }\mathbf{p}}+\sum_{\mathbf{pq}}\gamma _{\mu \mathbf{pq}%
}^{\ast }\gamma _{\mu ^{\prime }\mathbf{pq}}=\delta _{\mu \mu ^{\prime }}$,
are determined by solving the Schrodinger equation $H_{i}\left| \Psi _{i\mu
}\right\rangle =E_{\mu }\left| \Psi _{i\mu }\right\rangle $ \cite{note1},
where $\mu $ labels different eigenstates with the corresponding
eigen-energy $E_{\mu }$. This kind of two-particle equation has been solved
in Ref. \cite{7} with a Harmonic approximation to the potential well. Here,
we only need to mention two general features of the local dressed molecule
states: first, the eigen-energies $E_{\mu }$ of the dressed molecule can be
tuned by the external magnetic field $B$ through their dependence on $\nu
_{b}$, although such a dependence is in general nonlinear; second, the
energy difference between the adjacent eigenvalues $E_{\mu }$ is typically
of the order of the band gap energy $E_{bg}$.

We consider the case with one of the eigen-energies, say $E_{\mu _{0}}$,
pretty close to the two-atom free energy $2\epsilon _{\mathbf{p}%
_{0}}^{\left( a\right) }$ on a certain band $\mathbf{p}_{0}$,
i.e., we tune the magnetic field $B$ to satisfy the condition
$\left| E_{\mu _{0}}\left( B\right) -2\epsilon
_{\mathbf{p}_{0}}^{\left( a\right) }\right| \ll E_{bg}$, as
illustrated in Fig. 2. The $\mu _{0}$ and $\mathbf{p}_{0}$ can be
chosen respectively as the ground state of the dressed molecule
and the lattice lowest band, although they are not subject to such
a restriction. Under the above condition, if the atoms start in
the band $\mathbf{p}_{0}$, their state evolution will be
restricted in the Hilbert subspace involving only the excitations
of the modes $d_{i\mu _{0}}^{\dagger }$ and
$a_{i\mathbf{p}_{0}\sigma }^{\dagger }$, as all the other states
are significantly detuned by a energy scale of $\left(
E_{on},E_{bg}\right) $ which is much larger than $\left(
E_{off},E_{t}\right) $ \cite{note2}. So each site can only take
four possible states, given by $\left| 0\right\rangle $, $\left|
\sigma \right\rangle \equiv a_{i\mathbf{p}_{0}\sigma }^{\dagger
}\left|
0\right\rangle $ $\left( \sigma =\uparrow ,\downarrow \right) $,\ and $%
\left| d\right\rangle \equiv d_{i\mu _{0}}^{\dagger }\left| 0\right\rangle $%
. We can then project the full Hamiltonian $H=H_{0}+H_{I}$ into the physical
subspace specified by the projector $P\equiv \bigotimes_{i}P_{i}$, with $%
P_{i}\equiv \left| 0\right\rangle _{i}\left\langle 0\right| +\left| \uparrow
\right\rangle _{i}\left\langle \uparrow \right| +\left| \downarrow
\right\rangle _{i}\left\langle \downarrow \right| +\left| d\right\rangle
_{i}\left\langle d\right| $. After such a projection, the effective
Hamiltonian $H_{eff}\equiv PHP$ takes the form

\begin{figure}[tbph]
\centering
\includegraphics[height=3cm,width=8cm]{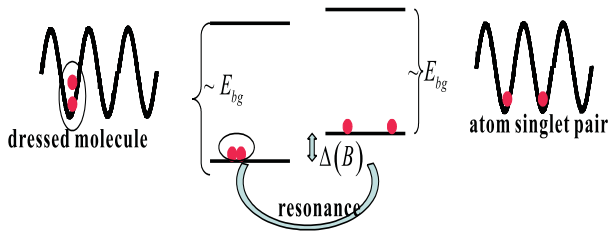}
\caption{ Illustration of resonance between the local dressed
molecules and the atom valence bonds on neighboring sites. }
\label{Fig2}
\end{figure}

\begin{widetext}

\begin{eqnarray}
H_{eff} &=&\sum_{i}\Delta \left( B\right) d_{i}^{\dagger
}d_{i}+\sum_{i}\sum_{j\in N\left( i\right) }\left[ \left( gd_{i}^{\dagger
}\left( a_{i\uparrow }a_{j\downarrow }-a_{i\downarrow }a_{j\uparrow }\right)
+h.c.\right) +t_{d}Pd_{i}^{\dagger }d_{j}P\right]  \nonumber \\
&&+\sum_{i}\sum_{j\in N\left( i\right) }\left[ t_{a}\sum_{\sigma
}Pa_{i\sigma }^{\dagger }a_{j\sigma }P+t_{da}\sum_{\sigma }d_{i}^{\dagger
}d_{j}a_{j\sigma }^{\dagger }a_{i\sigma }+c_{d}n_{id}n_{jd}+c_{a}\left(
n_{i}n_{j}/4-\mathbf{s}_{i}\cdot \mathbf{s}_{j}\right) \right] ,
\end{eqnarray}
where we have used the simplified notations $d_{i}\equiv d_{i\mu _{0}}$, $%
a_{i\sigma }=a_{i\mathbf{p}_{0}\sigma }$, $\Delta \left( B\right) \equiv
E_{\mu _{0}}\left( B\right) -2\epsilon _{\mathbf{p}_{0}}^{\left( a\right) }$
(the atom energy zero point \ has been shifted to coincide with $\epsilon _{%
\mathbf{p}_{0}}^{\left( a\right) }$), and defined the number and the spin
operators $n_{id}\equiv d_{i}^{\dagger }d_{i}$, $n_{i}\equiv \sum_{\sigma
}a_{i\sigma }^{\dagger }a_{i\sigma }$, $\mathbf{s}_{i}\equiv \sum_{\sigma
\sigma ^{\prime }}a_{i\sigma }^{\dagger }\mathbf{\sigma }_{\sigma \sigma
^{\prime }}a_{i\sigma ^{\prime }}/2$ with the Pauli matrix $\mathbf{\sigma =}%
\left( \sigma _{x},\sigma _{y},\sigma _{z}\right) $. The parameters in the
Hamiltonian $H_{eff}$ are given by $g=\gamma _{\mu _{0}\mathbf{p}_{0}\mathbf{%
p}_{0}}^{\ast }t_{\mathbf{p}_{0}}^{\left( a\right) }/2$ $+\sum_{\mathbf{q}%
}\chi _{\mu _{0}\mathbf{q}}^{\ast }c_{1;\mathbf{qp}_{0}\mathbf{p}%
_{0}}^{\left( am\right) }+\sum_{\mathbf{qs}}\gamma _{\mu _{0}\mathbf{qs}%
}^{\ast }c_{1;\mathbf{qsp}_{0}\mathbf{p}_{0}}^{\left( aa\right) }$, $%
t_{d}=\sum_{\mathbf{q}}\left| \chi _{\mu _{0}\mathbf{q}}\right| ^{2}t_{%
\mathbf{q}}^{\left( m\right) }$ $+2\mathop{\rm Re}\nolimits(\sum_{\mathbf{qss%
}^{\prime }}\chi _{\mu _{0}\mathbf{q}}^{\ast }c_{2;\mathbf{qss}^{\prime
}}^{\left( am\right) }\gamma _{\mu _{0}\mathbf{s}^{\prime }\mathbf{s}%
})+\sum_{\mathbf{pqss}^{\prime }}\gamma _{\mu _{0}\mathbf{pq}}^{\ast }c_{2;%
\mathbf{pqss}^{\prime }}^{\left( aa\right) }\gamma _{\mu _{0}\mathbf{s}%
^{\prime }\mathbf{s}}$, $t_{a}=t_{\mathbf{p}_{0}}^{\left( a\right) }$, $%
t_{da}=-\sum_{\mathbf{q}}\left| \gamma _{\mu _{0}\mathbf{qp}_{0}}\right|
^{2}t_{\mathbf{q}}^{\left( a\right) }$ $-2\mathop{\rm Re}\nolimits(\sum_{%
\mathbf{qs}}\chi _{\mu _{0}\mathbf{q}}^{\ast }c_{1;\mathbf{qp}_{0}\mathbf{s}%
}^{\left( am\right) }\gamma _{\mu _{0}\mathbf{sp}_{0}})-2\mathop{\rm Re}%
\nolimits(\sum_{\mathbf{pqs}}\gamma _{\mu _{0}\mathbf{pq}}^{\ast }c_{1;%
\mathbf{pqp}_{0}\mathbf{s}}^{\left( aa\right) }\gamma _{\mu _{0}\mathbf{sp}%
_{0}})$,\ $c_{d}=\sum_{\mathbf{pp}^{\prime }\mathbf{qq}^{\prime }\mathbf{ss}%
^{\prime }}\gamma _{\mu _{0}\mathbf{pp}^{\prime }}^{\ast }\gamma _{\mu _{0}%
\mathbf{q}^{\prime }\mathbf{q}}^{\ast }c_{3;\mathbf{pqss}^{\prime }}^{\left(
aa\right) }\gamma _{\mu _{0}\mathbf{sp}^{\prime }}\gamma _{\mu _{0}\mathbf{q}%
^{\prime }\mathbf{s}^{\prime }}$, $c_{a}=c_{3;\mathbf{p}_{0}\mathbf{p}_{0}%
\mathbf{p}_{0}\mathbf{p}_{0}}^{\left( aa\right) }$.

\end{widetext}

The Hamiltonian $H_{eff}$ represents an effective single-band model, but it
has incorporated multi-band information into its parameters through the
structure coefficients $\chi _{\mu _{0}\mathbf{p}}$,$\gamma _{\mu _{0}%
\mathbf{pq}}$ of the local dressed molecules. This Hamiltonian
contains no on-site interactions as they have been exactly taken
into account into the structure of the dressed molecules. As one
tunes $\Delta \left( B\right) $ through the magnetic field, the
Hamiltonian $H_{eff}$ describes a crossing resonance between the
local dressed molecules $d_{i}$ and the atomic valence bonds
$a_{i\uparrow }a_{j\downarrow }-a_{i\downarrow }a_{j\uparrow }$ on
the neighboring sites. This resonance may have a richer structure
than the conventional Feshbach resonance between the bare
molecules and the free atoms as first, the dressed molecule here
has been a complicated superposition of the bare molecules and the
local Cooper pairs, and second, the resonance to valence bonds on
neighboring sites may introduce rich configurations depending on
the lattice geometry.

The Hamiltonian $H_{eff}$ supports rich physics. Its detailed
study will be presented elsewhere. Here, we investigate $H_{eff}$
in two limiting cases with the detuning $\left| \Delta \left(
B\right) \right| $ significantly larger than the neighboring
coupling rate $g$. We can see already rich phase diagrams from
$H_{eff}$ in these limiting cases. First, we consider the case
with the population dominantly in the atoms and the local dressed
molecules only virtually excited due to the large detuning $\Delta
\left( B\right) $ (the lattice filling number $\overline{n}\leq 1$
in this case). We can then reduce $H_{eff}$ to an effective
Hamiltonian involving only the atomic operators $a_{i\sigma }$.
For this purpose, we define the projection
operator $P_{a}\equiv \bigotimes_{i}P_{i}^{\left( a\right) }$ with $%
P_{i}^{\left( a\right) }\equiv \left| 0\right\rangle _{i}\left\langle
0\right| +\left| \uparrow \right\rangle _{i}\left\langle \uparrow \right|
+\left| \downarrow \right\rangle _{i}\left\langle \downarrow \right| $, and
keep the terms up to the order of $\left| g\right| ^{2}/\Delta \left(
B\right) $ in the projection. The reduced Hamiltonian can be derived through
either the second-order perturbation or a canonical transformation \cite{11}%
, and it takes the form of the famous $t$-$J$ model

\begin{equation}
H_{tJ}=\sum_{i;j\in N\left( i\right) }\left[ t_{a}\sum_{\sigma
}P_{a}a_{i\sigma }^{\dagger }a_{j\sigma }P_{a}+J\left( \mathbf{s}_{i}\cdot
\mathbf{s}_{j}-n_{i}n_{j}/4\right) \right] ,
\end{equation}
where the parameter $J=-c_{a}-2\left| g\right| ^{2}/\Delta \left(
B\right) $ \cite{note3}. In $H_{tJ}$, we have not included the
$3$-site nonlinear tunnelling $H_{3}=\left[ -\left| g\right|
^{2}/\Delta \left( B\right) \right] \sum_{i}\sum_{j,k\in N\left(
i\right) }\left( a_{i\uparrow }^{\dagger }a_{k\downarrow
}^{\dagger }-a_{i\downarrow }^{\dagger }a_{k\uparrow }^{\dagger
}\right) \times \left( a_{i\uparrow }a_{j\downarrow
}-a_{i\downarrow }a_{j\uparrow }\right) $, which is usually
omitted in the literature \cite{note4}. The $t$-$J$ model plays a
fundamental role in study
of high $T_{c}$ superconductivity \cite{10,11,12}. Depending on the ratio $%
t_{a}/J$, the atom filling $\overline{n}$, and the lattice dimension and
geometry, the $t$-$J$ model shows rich phase diagrams, including, for
instance, the atomic anti-ferromagnetic phase, the $d$-wave
superconductivity \cite{14}, and likely the resonating valence bond (RVB)
states \cite{10,11}. The $t$-$J$ model was previously derived as the strong
interaction limit of the Hubbard model \cite{10,12}. Here, although our
basic Hamiltonian $H_{eff}$ does not resemble any forms of the Hubbard
model, we get exactly the same reduced Hamiltonian as the $t$-$J$ model;
furthermore, our atomic realization of the $t$-$J$ model is not subject to
the constraint $J\ll t_{a}$ as is the case from the Hubbard model. All the
parameters here can be experimentally tuned in the full range, which,
together with the easy control of the lattice dimension and geometry, makes
this system the ideal test bed for many outstanding predictions and
hypotheses associated with the $t$-$J$ model.

We now consider another limiting case of our basic Hamiltonian $H_{eff}$
with the population dominantly in the dressed molecules and the atoms only
virtually excited (still, $\left| \Delta \left( B\right) \right| \gg \left|
g\right| $). In this case, we define the molecule projection operator $%
P_{m}\equiv \bigotimes_{i}P_{i}^{\left( m\right) }$ with $P_{i}^{\left(
m\right) }\equiv \left| 0\right\rangle _{i}\left\langle 0\right| +\left|
d\right\rangle _{i}\left\langle d\right| $. Up to the order of $\left|
g\right| ^{2}/\Delta \left( B\right) $, the reduced Hamiltonian for the
dressed molecules has the form
\begin{equation}
H_{m}=\sum_{i}\left[ \Delta ^{\prime }n_{id}+\sum_{j\in N\left( i\right)
}\left( t_{d}^{\prime }P_{m}d_{i}^{\dagger }d_{j}P_{m}+c_{d}^{\prime
}n_{id}n_{jd}\right) \right]
\end{equation}
where $t_{d}^{\prime }=t_{d}+2\left| g\right| ^{2}/\Delta \left( B\right) ,$
$c_{d}^{\prime }=c_{d}-2\left| g\right| ^{2}/\Delta \left( B\right) $, and $%
\Delta ^{\prime }=\Delta \left( B\right) +2n_{c}\left| g\right| ^{2}/\Delta
\left( B\right) $ ($n_{c}$ is the lattice coordination number). This
Hamiltonian describes hard-core bosons with neighboring interactions, and it
is identical to the magnetic XXZ model $H_{XXZ}=\left( t_{d}^{\prime
}/4\right) \sum_{i}\left[ B_{eff}Z_{i}+\sum_{j\in N\left( i\right) }\left(
X_{i}X_{j}+Y_{i}Y_{j}+\delta _{z}Z_{i}Z_{j}\right) \right] ,$ $\left( \delta
_{z}=c_{d}^{\prime }/t_{d}^{\prime }\text{, }B_{eff}=2\left( \Delta ^{\prime
}+n_{c}c_{d}^{\prime }\right) /t_{d}^{\prime }\right) $\ for the effective
Pauli operators $X_{i},Y_{i},Z_{i}$ through the following mapping: $%
X_{i}\equiv P_{m}\left( d_{i}+d_{i}^{\dagger }\right) P_{m}$,
$Y_{i}\equiv iP_{m}\left( d_{i}-d_{i}^{\dagger }\right) P_{m}$,
$Z_{i}\equiv 2d_{i}^{\dagger }d_{i}-1$ \cite{14,6}. The phase
diagram of the XXZ model is know pretty well. For instance, it has
the ferromagnetic, the canted XY, and the anti-ferromagnetic
phases \cite{15}, which correspond respectively to the Mott state,
the superfluidity state, and the checkerboard state (one
occupation every other sites) of the dressed molecules. It also
has more exotic quantum phases (RVB\ spin liquids etc.) if the
lattice geometry induces frustration of the above spin orders
\cite{16}.

In summary, we have derived the effective Hamiltonian for
fermionic atoms in an optical lattice across broad Feshbach
resonance, taking into account of both the multi-band
configurations and the direct neighboring interactions. In certain
limits, this Hamiltonian is reduced to either the $t$-$J$ model
for the fermionic atoms or the XXZ model for the local dressed
molecules. The latter two models connect to many fundamental
physics associated with many-body systems.

This work was supported by the NSF awards (0431476), the ARDA under ARO
contracts, and the A. P. Sloan Fellowship.


\begin{thebibliography}{99}
\bibitem{note1}  In solving this two-atom Schrodinger equation, one needs to
apply some renormalization procedure as has been specified in Ref. \cite{7}.
\bibitem{note2}  These other states include $3$ or more atom occupations of
a single site, which take an energy cost typically of the order of $E_{on}$,
and thus will not be excited in our considered case.
\bibitem{note3}  In this expression of $J$, we have included the virtual
excitation contribution from only the nearest dressed molecule level. If the
detuning $\Delta \left( B\right) $ becomes comparable with the band gap $%
E_{bg}$, one should include in $J$ the virtual excitation contributions from
several nearby dressed molecule levels, but the reduced Hamiltonian for the
atoms still takes the form of the $t$-$J$ model.
\bibitem{note4}  This omission is based on the believe that the main physics
of the $t$-$J$ model is caught by the competition between the $t$ and $J$
terms, and the $3$-site tunneling is not as important as the $t$ term, see,
e.g., the discussion on page 27 in Ref. \cite{11}.
\end{thebibliography}
\end{document}